\documentclass[floats,floatfix,showpacs,amssymb,prd,twocolumn,superscriptaddress,nofootinbib]{revtex4-2}

\usepackage{graphicx}
\usepackage{dcolumn}
\usepackage{bm}
\usepackage{amsmath}
\usepackage[colorlinks=true,linkcolor=blue,citecolor=blue,urlcolor=blue]{hyperref}
\usepackage{dsfont}
\usepackage{xcolor}

\begin{document}

\title{Efficient numerical evolution of self-gravitating systems\\ with dark-matter--baryon scattering}

\author{Marc Kamionkowski}
\email{kamion@jhu.edu}
\affiliation{William H.\ Miller III Department of Physics \& Astronomy, Johns Hopkins University, 3400 N.\ Charles St., Baltimore, MD 21218, USA}

\author{Vera Gluscevic}
\affiliation{Department of Physics \& Astronomy, University of Southern California, Los Angeles, CA 90089, USA}
\affiliation{Institute for Advanced Study, 1 Einstein Drive, Princeton, NJ 08540, USA}
\affiliation{Center for Computational Astrophysics, Flatiron Institute, 162 5th Avenue, New York, NY 10010, USA}

\author{Kris Sigurdson}
\affiliation{Department of Physics and Astronomy, University of British Columbia, Vancouver, BC V6T 1Z1, Canada}

\date{\today}

\begin{abstract}
We present a fast algorithm to numerically evolve a spherically
symmetric system consisting of dark matter and gas, including dark-matter--baryon scattering.  The method augments the spherical $N$-body code {\tt NSphere}, which collapses the
six-dimensional phase space of a spherical collisionless system to
three dimensions, with a one-dimensional Lagrangian treatment of gas.  Dark-matter--baryon scattering is implemented as a Monte-Carlo process between the dark-matter numerical particles and the gas shells.  We also implement an algorithm for radiative gas cooling that is valid for weak cooling.  The gas, as well as dark-matter--baryon scattering and cooling, adds only incrementally to the run time of the code, allowing systems to be evolved on a laptop computer, a tiny fraction of the computational resources required for a full  3D $N$-body/hydro simulation.   To illustrate, we show the evolution of a $\sim 10^{13}\,M_\odot$ halo with a dark-matter--baryon scattering cross section $\sigma_{\rm tot}/m_\chi=0.1\,{\rm cm}^2\,{\rm g}^{-1}$ but no gas cooling.  We then show examples with radiative gas cooling where the cooling is altered through interaction with dark-matter particles.  The resulting code, {\tt NSphere-IDM}, allows the implications of dark-matter interactions for galaxy evolution to be explored quickly over large volumes of the dark-matter parameter space, and other new physics (e.g., dark-matter self-interactions or anisotropy in the distribution function) is easily incorporated.  The code can also provide useful calibrations and benchmarks for fully  3D $N$-body or hydrodynamical simulations.
\end{abstract}

\maketitle

\section{Introduction}
\label{sec:intro}

A broad effort has gone into using  astrophysical and cosmological observations to constrain and characterize
dark-matter interactions with baryons \cite{Buen-Abad:2021mvc,Gluscevic:2019yal,Boddy:2022knd}.  Much of this work has been within the context of specific dark-matter models, such as dipole \cite{Sigurdson:2004zp} or millicharged \cite{Davidson:2000hf, Kovetz:2018zan} dark matter, but there is also a broader effort to study such interactions in a model-independent way \cite{Starkman:1990nj,Boehm:2004th,Dvorkin:2013cea,Munoz:2015bca,Ali-Haimoud:2015pwa,Kovetz:2018zan,Gluscevic:2017ywp,Slatyer:2018aqg,Xu:2018efh,Boddy:2018wzy,Buen-Abad:2021mvc,Maamari:2020aqz,Nadler:2019zrb, DES:2020fxi,He:2023dbn,He:2025npy, Crumrine:2024sdn,Crumrine:2026rnp,An:2024nsw, Dvorkin:2020xga,Nguyen:2021cnb}. At early times such interacting dark matter (IDM) damps small-scale structure, and the resulting bounds from
the cosmic microwave background (CMB) and/or large-scale structure \cite{Dvorkin:2013cea,Gluscevic:2017ywp,AtacamaCosmologyTelescope:2025nti}, Milky-Way satellite
population \cite{Nadler:2019zrb,DES:2020fxi,Maamari:2020aqz}, and the Lyman-$\alpha$ forest measurements~\cite{Rogers:2021byl} are now stringent.  Far less is known about the effect of dark-matter--baryon scattering on
individual halos, where gas can condense in the potential well and the relative bulk velocities between the dark matter and gas can again make scattering efficient.  $N$-body/hydro simulations of dark matter and gas are required to understand this regime,  and the first simulations that explicitly implement dark-matter--baryon scattering have only recently appeared \cite{Fischer:2025snw,Wen:2026,Hainje:2026qou}.

These simulations are expensive: they evolve a fully three-dimensional system with $N$-body gravity and a hydrodynamic solver, and the space of IDM models, including cross-section amplitude and velocity dependence, dark-matter mass, and halo mass, is large (and even larger considering that models with IDM also often imply dark-matter self-interactions).  Analytic predictions to benchmark the numerical implementation of new dark-matter physics in the context of IDM are also lacking. It is therefore useful to have a fast,
complementary tool that can survey this space and develop physical
intuition, even at the cost of simplifying assumptions.  

Here we extend {\tt NSphere}, a code that evolves
spherically symmetric self-gravitating systems of collisionless
particles~\cite{Kamionkowski:2025uae} (or particles with self-interactions \cite{Kamionkowski:2025fat}---see also Ref.~\cite{Gurian:2025zpc} for a similar approach), to include a gas component and dark-matter--baryon scattering.  The economy of {\tt NSphere} comes from collapsing the full six-dimensional phase space to three
dimensions (radius, speed, and the angle the velocity makes with the
radial direction) and determining the force on any given particle from Newton's shell theorem instead of calculating  pairwise forces for all numerical particles, with a small correction for shell crossing each time step.  We add to this picture a gas component, evolved as a set of concentric Lagrangian shells, and a Monte-Carlo
treatment of dark-matter--baryon scattering similar to that used for self-interacting dark matter (SIDM) \cite{Kamionkowski:2025fat, Rocha:2012jg}.  We also incorporate an algorithm to treat radiative cooling that is valid when the cooling time is long compared with the dynamical time.\footnote{Examples include the intracluster medium in galaxy clusters and gas in massive halos, especially early in their evolution.}  We show how to initialize a static system of dark matter with gas in hydrostatic equilibrium so that the effects of cooling, dark-matter--baryon scattering, and/or self-interactions, can be isolated.  The addition of the fluid increases the run time of {\tt  NSphere} only marginally, allowing full simulations to be evolved (with or without dark-matter--baryon scattering and/or cooling) on a laptop computer.  We provide the code, {\tt NSphere-IDM}, at \url{https://github.com/NSphere-IDM/NSphere-IDM} to allow reproduction of our results and as a starting point for those who may wish to develop this approach further.

The paper is organized as follows.
Section~\ref{sec:initialconditions} describes initial conditions for a self-consistent dark-matter distribution function with gas in hydrostatic equilibrium.  Section~\ref{sec:gas} provides the Lagrangian equations for the gas.  Section~\ref{sec:scattering} describes the Monte-Carlo algorithm for dark-matter--baryon scattering and how the resulting exchange of energy and momentum between the gas and dark matter is implemented.  Section~\ref{sec:cooling}
describes how we incorporate radiative cooling.
Section~\ref{sec:application} shows the evolution of the gas in a $10^{13}\, M_\odot$ galaxy in the presence of dark-matter--baryon scattering, and Section~\ref{sec:coolingdemo} shows how cooling flows in clusters can be slowed (or accelerated) by dark-matter--baryon scattering. Section~\ref{sec:conclusions} discusses some of the limitations of this approach but also some expansions that should be easy to add.  Appendix~\ref{app:gas} provides additional details on the numerical implementation of the gas equations. Appendix~\ref{app:sigmaeff} derives the effective cross section that maps our free-nucleon scattering model onto a hydrogen--helium or metal-rich gas.

\section{Initial conditions}
\label{sec:initialconditions}

\subsection{The self-consistent distribution function}

We first review the construction of a self-consistent spherically
symmetric self-gravitating system of particles of density $\rho(r)$,
with $\rho(r)\to0$ as $r\to\infty$ and finite total mass
$M_h=M(R)\equiv4\pi\int_0^R dr\,r^2\rho(r)$.  The relative potential is
$\Psi(r)=-\Phi(r)+\Phi_0$, with $\Phi(r)$ the usual gravitational
potential, so that $\nabla^2\Psi=-4\pi G\rho$ with $\Psi(r)\to0$ as
$r\to\infty$; its radial gradient is $\partial\Psi/\partial r=-GM(r)/r^2$.
A particle of speed $v$ has relative energy ${\cal E}=\Psi(r)-\tfrac12v^2$.

If the distribution function (DF) depends only on energy, $f({\cal E})$, it is obtained from the
density profile through Eddington's formula \cite{Eddington:1916},
\begin{equation}
     f({\cal E}) = \frac{1}{m_\chi\sqrt{8}\pi^2}
     \frac{d}{d{\cal E}}\int_0^{\cal E}
     \frac{d\Psi}{\sqrt{{\cal E}-\Psi}}\,\frac{d\rho}{d\Psi},
\label{eqn:DF}
\end{equation}
with $m_\chi$ the dark-matter particle mass.  A DF that depends on energy ${\cal E}$ only (and not on the angular momentum $L$) gives an isotropic velocity distribution everywhere.

Adding gas with mass density $\rho_{\rm gas}(r)$, the
total density becomes $\rho=\rho_{\rm dm}+\rho_{\rm gas}$ and the potential is sourced by the total density, $\nabla^2\Psi=-4\pi G\rho$.  Because Eq.~(\ref{eqn:DF}) is linear in the density at a  fixed total potential $\Psi$, the dark-matter DF is obtained simply by replacing $\rho$ in Eq.~(\ref{eqn:DF}) with $\rho_{\rm dm}$, the dark-matter density.
This is the only modification to the dark-matter initial conditions: the
$r(\Psi)$ map and the gradient $d\Psi/dr=-GM_{\rm tot}(<r)/r^2$ that enter Eq.~(\ref{eqn:DF}) both use the {\it total} enclosed mass, so the sampled dark matter is in equilibrium in the combined potential.\footnote{Similar augmentations are discussed in Refs.~\cite{Barnea,Posner}.}

More generally, self-consistent DFs for spherically symmetric systems are functions $F({\cal E},L)$ of the particle angular momentum $L$ as well, thus allowing for DFs with anisotropic velocity distributions.  These DFs are discussed in Ref.~\cite{Kamionkowski:2025syj}.  We note that our current code also allows for anisotropic DFs with IDM, although they are not discussed further in this paper.

\subsection{Generating the realization}

The realization of the dark-matter phase-space DF is generated exactly as in Ref.~\cite{Kamionkowski:2025uae}: a radius is drawn from the cumulative mass profile $M_{\rm dm}(<r)/M_{\rm dm}$, a speed is drawn by rejection sampling from $v^2 f[\Psi(r)-v^2/2]$, and a direction cosine $\mu=\cos\theta$ is drawn uniformly on $[-1,1]$ (or from the appropriate nonuniform distribution if the dependence of the DF on velocity direction is nontrivial).  The only change is that $\Psi(r)$ now includes the gas contribution.

\section{Equations for the gas}
\label{sec:gas}

We treat the gas as a monatomic ideal fluid with equation of state
$P=(\rho_{\rm gas}/\mu m_H)k_B T$ and specific internal energy
$u=(3/2)k_B T/(\mu m_H)$, so that $P=(2/3)\rho_{\rm gas}u$, with $\mu$ the mean molecular weight ($\mu=1$ for neutral atomic hydrogen); $k_B$ is the Boltzmann constant.  Using the enclosed gas mass $m$ as the independent (Lagrangian) variable, gas-mass conservation is $\partial r/\partial m=(4\pi r^2\rho_{\rm gas})^{-1}$, and the momentum equation is
\begin{equation}
     \frac{\partial^2 r}{\partial t^2} = -\frac{G M_{\rm tot}(<r)}{r^2}
     - 4\pi r^2 \frac{\partial P}{\partial m} + F_{\chi b},
\label{eqn:fluidacceleration}
\end{equation}
where $M_{\rm tot}(<r)$ includes both dark matter and gas, and $F_{\chi b}$ is the specific (per-unit-mass) radial force imparted to the gas by dark-matter--baryon scattering.  The internal energy of gas obeys the first law of thermodynamics,
\begin{equation}
     \frac{\partial u}{\partial t} = \frac{P}{\rho_{\rm gas}^2}\frac{\partial\rho_{\rm gas}}
     {\partial t} + \epsilon,
\label{eqn:energy}
\end{equation}
with $\epsilon$ the specific rate at which energy is deposited to (or, if $\epsilon<0$, removed from) the gas.  The evolution is thus governed by three first-order equations for $r(m;t)$, $v_r(m;t)$, and $T(m;t)$, with $\partial\rho_{\rm gas}/\partial t$ obtained from
$\rho_{\rm gas}^{-1}\partial\rho_{\rm gas}/\partial t=-2 (v_r/r)-\partial_t(\partial r/\partial m)/(\partial r/\partial m)$
and $\partial_t(\partial r/\partial m)=\partial v_r/\partial m$.

The equations are discretized on $N_{\rm gas}$ equal-gas-mass
shells with edges $r_i$, edge velocities $v_i$, and shell-centered
$\rho_{i+\frac12}$, $P_{i+\frac12}$, $u_{i+\frac12}$, advanced by a
kick--drift--kick leapfrog.

The gas is initialized in hydrostatic equilibrium, $dP/dr=-GM_{\rm tot}(<r)\rho_{\rm gas}(r)/r^2$, for a chosen dark-matter profile $\rho_{\rm dm}(r)$ and gas profile $\rho_{\rm gas}(r)$, subject to $P(r\to\infty)=0$; the initial temperature follows from the equation of state.  

\section{Dark-matter--baryon scattering}
\label{sec:scattering}

Scattering of dark-matter particles from the gas is implemented as a Monte-Carlo process. In each step, for each dark-matter numerical particle we do the following: (a) compute the probability that it scatters off a gas atom (or ion) in the gas shell it occupies; (b) if it does, draw a partner atom and determine a final dark-matter velocity and direction; and (c) deposit the resulting  energy change and the radial component of the momentum change, with the opposite sign, into the gas shell, where they enter Eqs.~(\ref{eqn:fluidacceleration}) and (\ref{eqn:energy}) as $F_{\chi b}$ and a contribution $\epsilon_{\chi b}$ to $\epsilon$.

\subsection{Kinematics of dark-matter--baryon scattering}

A dark-matter particle at radius $r$ with speed $v_{\chi1}$ and direction cosine
$\mu_1$ in the lab frame moving through gas of density $\rho_{\rm gas}$, temperature $T$, and bulk
radial velocity $v_{\rm gas}$ scatters in a time interval $\Delta t$ with probability  $n\langle \sigma_{\rm tot}v_{\rm rel}\rangle\Delta t$, where $n=\rho_{\rm gas}/m_H$, and the average is over the atom's thermal velocity distribution.

This scattering rate uses the total baryon number density, $n=\rho_{\rm gas}/m_H$, with a single cross section $\sigma_{\rm tot}$ per baryon and free-baryon kinematics, which captures the bulk thermodynamics and is representative of plausible microphysics: for $m_\chi\ll m_H$ it reproduces the energy-exchange rate of coherent scattering with equal proton and neutron couplings to $\sim$5--10\%, and more generally any coherent coupling to protons and neutrons maps onto an effective cross section $\sigma_{\rm tot}\rightarrow\sigma_{\rm tot}^{\rm eff}$; for more details see Appendix~\ref{app:sigmaeff}.

In the gas rest frame the dark-matter speed is
$w=(v_{\chi1}^2-2\mu_1 v_{\chi1}v_{\rm gas}+v_{\rm gas}^2)^{1/2}$ and the
atoms have a Maxwell-Boltzmann velocity distribution of width set by $\beta\equiv m_H/(2k_BT)$. Averaging over the atom velocities reduces the rate to a one-dimensional integral over the relative speed; for a constant cross section it is
analytic,
\begin{equation}
\langle\sigma_{\rm tot} v_{\rm rel}\rangle
= \sigma_{\rm tot}\!\left[
  \left(w+\frac{1}{2\beta w}\right)\!\mathrm{erf}\!\left(\sqrt{\beta}\,w\right)
  + \frac{e^{-\beta w^2}}{\sqrt{\pi\beta}}\right].
\label{eqn:sigmav}
\end{equation}
For other cross sections the integral can be pre-tabulated in $(w,\beta)$.  Note that in this formulation, the gas {\it thermal} velocities are retained in the rate through $\beta$.  If a randomly drawn number between 0 and 1 is smaller than the probability   $n\langle \sigma_{\rm tot}v_{\rm rel}\rangle\Delta t$, then the particle scatters.\footnote{More precisely, we use the Poisson probability of at least one scattering, $1-\exp(-n\langle \sigma_{\rm tot}v_{\rm rel}\rangle\Delta t)$, which coincides with the linear form for our small time steps.}

If the particle scatters, the next step is to determine which particle it scatters from.  To do so, we choose the atom from a velocity ${\bf u}$ distribution $f({\bf u}) \sigma_{\rm tot}(v_{\rm rel}) v_{\rm rel}$ by first choosing the three components of $\bf u$ from a Gaussian of width $(k_B T/m_H)^{1/2}$.  We then choose a random number between 0 and 1 and accept the atom if that number is less than $\sigma_{\rm tot}(v_{\rm rel}) v_{\rm rel}/\left[\sigma_{\rm tot}(v_{\rm rel}) v_{\rm rel}\right]_{\rm max}$, where $\left[\sigma_{\rm tot}(v_{\rm rel}) v_{\rm rel}\right]_{\rm max}$ is a suitable upper bound.  If not, we repeat until we find an atom.

Once we have determined the ${\bf u}$ of the scattering atom, we need to determine the final dark-matter velocity $v_{\chi2}$ and direction $\mu_2$.  We first find the center-of-mass velocity ${\bf v}_{\rm cm}=({\bf v}_{\chi1} m_\chi+{\bf v}_{H} m_H)/(m_\chi+m_H)$ which can be used to boost in and out of the CoM frame (here ${\bf u} + {\hat r} v_{\rm gas}$ is the atom velocity).  The direction of the initial relative velocity (which is frame-independent) is $\hat v_{\rm rel}$.  Since we are assuming elastic scattering, the outgoing CoM velocities of the  dark-matter particle and atom have the same magnitudes (in the CoM frame) as the initial velocities and are oriented in opposite directions with angles $\alpha$ with respect to $\hat v_{\rm rel}$ that are chosen from $d\sigma/d\alpha$.  For example, if the scattering is isotropic, this implies that  $\cos\alpha$ is uniformly distributed between $-1$ and 1.  Additionally,  an azimuthal angle $\phi_f$ around $\hat v_{\rm rel}$ is chosen from a uniform distribution between 0 and $2\pi$.  These angles fully define the outgoing CoM velocity of the dark-matter particle which is then boosted back to the lab frame to give us ${\bf v}_{\chi2}$ which is then projected onto the $\hat r$ axis to give us $\mu_2$.  In this way, we obtain the final dark-matter particle velocity ${\bf v}_{\chi2}$ from the initial velocity ${\bf v}_{\chi1}$.

\subsection{Energy/momentum deposition in gas}

The change in the dark-matter numerical particle's momentum contributes to $F_{\chi b}$, the gas acceleration.  The change in the energy of the dark-matter numerical particle goes into $\epsilon$, the heat source/sink for the gas, and to a lesser extent, the kinetic energy of the gas.  The dark matter is represented in the simulation by numerical particles of mass  $m_p=M_{\rm dm}/N$ and the gas by shells of mass $\Delta m_{\rm gas} =M_{\rm gas}/N_{\rm gas}$.  The change in the energy of the numerical particle is $\Delta E=m_p (v_{\chi2}^2-v_{\chi1}^2)/2$, and the change to its momentum is $\Delta p=m_p(v_{\chi2}\mu_2-v_{\chi1}\mu_1)\equiv m_p\Delta v_r$.  The dark-matter particle scatters from an individual atom, but the energy and momentum changes to the atom are then distributed to the gas shell as a whole. Only the radial momentum acts on a spherical shell: the tangential kicks point in all directions and cancel over many scatters.  The kinetic energy induced in the gas shell by the radial-momentum change is thus $(\Delta p)^2/(2 \Delta m_{\rm gas})$.  The remainder of the energy goes into the gas, but the heat is determined in the rest frame of the gas where the shell starts at rest and has a recoil
energy $\Delta p^2/(2\Delta m_{\rm gas})$.  We then obtain a contribution
\begin{equation}
  \Delta U=-\frac12 m_p\,(w_{\chi2}^2-w_{\chi1}^2)-\frac{\Delta p^2}{2\Delta m_{\rm gas}},
\label{eqn:deposit}
\end{equation}
where $w_{\chi1}$ and $w_{\chi2}$ are respectively the initial and final dark-matter speed in the gas frame.  In lab variables this becomes
\begin{equation}
     \Delta U=-\frac12 m_p(v_{\chi2}^2-v_{\chi1}^2)+v_{\rm gas}\,\Delta p-\frac{\Delta p^2}{2\Delta m_{\rm gas}}.
\label{eq:lab}
\end{equation}
The cross term  $v_{\rm gas}\Delta p$ is the work the kick does on the shell's pre-existing bulk motion.

The contribution from this scatter to the gas acceleration $F_{\chi b}$ and gas-heating rate $\epsilon_{\chi b}$ is then $-(\Delta p) \,/(\Delta m_{\rm gas}\,\Delta t)$ and $\Delta U /(\Delta m_{\rm gas} \Delta t)$, where $\Delta t$ is the time step.

\begin{figure*}[!t]
\includegraphics[width=\textwidth]{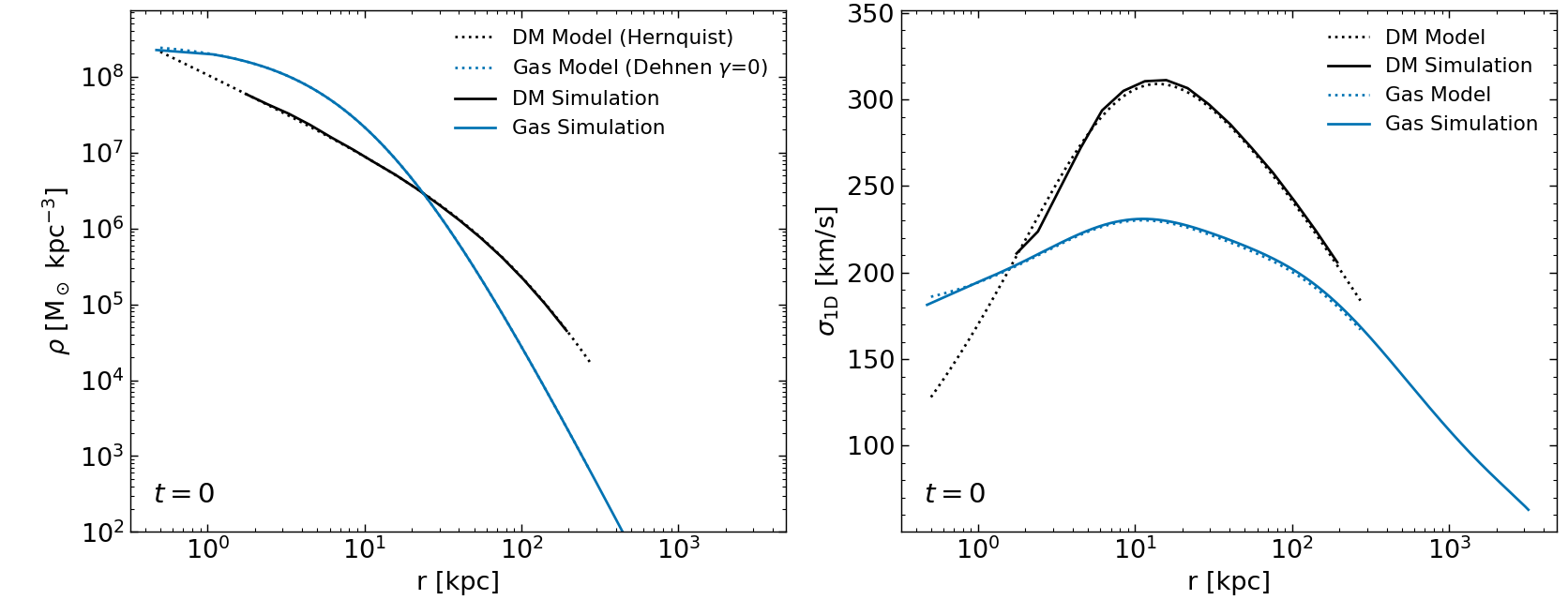}
\caption{{\it Cool-core run.}  The self-consistent initial conditions ($t=0$).  {\it Left:}
spherically averaged density of the dark matter (black) and gas (blue) from the sampled initial conditions (solid), with the analytic Hernquist and
Dehnen models (dotted).  The gas is more centrally concentrated
and dominates inside $\sim20$~kpc.  {\it Right:} the one-dimensional velocity
dispersion of the dark matter  (black) and the one-dimensional gas thermal dispersion $\sigma_{\rm gas}=
\sqrt{(2/3)u}$  (blue) from the same initial conditions (solid), with the
analytic isotropic-Jeans prediction for the dark matter and the hydrostatic-equilibrium
prediction for the gas (dotted).  Both models are evaluated in the combined
dark-matter-plus-gas potential.  The sampled profiles reproduce the analytic models over two
decades in radius, for both components.  }
\label{fig:ics}
\end{figure*}

\section{Radiative cooling}
\label{sec:cooling}

Radiative cooling is added as a local sink in Eq.~(\ref{eqn:energy}).  We write
$\epsilon=\epsilon_{\chi b}+\epsilon_{\rm cool}$ with
\begin{equation}
\epsilon_{\rm cool} = -\frac{n_e n_t}{\rho_{\rm gas}}\Lambda(T,Z)
= -\,\left[1+\frac{Y}{4X}\right]\frac{x_e X^2 \rho_{\rm gas}}{m_H^2}\,\Lambda(T,Z),
\label{eqn:epscool}
\end{equation}
where $\Lambda(T,Z)$ is the cooling function (which we obtain from
collisional-ionization-equilibrium calculations~\cite{Sutherland:1993ec,Wiersma:2008cs}), $X$ ($=0.71$) is the hydrogen mass fraction, and $x_e=n_e/n_H$, with $n_H=X\rho_{\rm gas}/m_H$ the hydrogen number density. We use the normalization of Ref.~\cite{Sutherland:1993ec}, in which the volumetric loss is $n_e n_t \Lambda$ with $n_t=[1+Y/(4X)]\,n_H\simeq1.1\,n_H$, where $Y=0.27$ is the helium mass fraction.  
Since $\Lambda$ is tabulated, cooling is integrated with the exact, unconditionally stable scheme of Ref.~\cite{Townsend:2009mf}.  

Since we are not (yet) attempting to model the transition of gas into stars, we impose a temperature floor of $10^4$~K.  Also our Lagrangian treatment of the gas cannot handle shocks, and so we cannot evolve the system into regimes where the dynamical time is longer than the cooling time $t_{\rm cool}=u/|\epsilon_{\rm cool}|$ and pressure support is lost.  We do, however, still add safeguards to keep the Lagrangian shells well ordered as radiative cooling sharpens the pressure gradients.  First, after each drift of the shell edges, $r_i \to r_i + v_i\,\Delta t$, the edges are clamped onto a monotonic
grid: the innermost edge is floored at $r_0\ge0$, and every edge is required
to exceed its inner neighbor by at least a minimum thickness,
$r_i \ge r_{i-1} + \delta r$ with $\delta r = 10^{-4}\,a_{\rm gas}$ (where $a_{\rm gas}$ is the gas scale radius defined below); an edge that violates this (or that returns a non-finite value) is reset to $r_{i-1}+\delta r$. This prevents adjacent shells from crossing or collapsing through the origin. Second, the global step is advanced by Courant-limited subcycling, with substep $\Delta t_{\rm gas} = C_{\rm cfl}\,\min_i\!\big[\Delta r_i/(c_{s,i}+|v_i|)\big]$
and $C_{\rm cfl}=0.3$, so that an edge advances by only a fraction of its
shell width per substep; shells already pinned at the floor thickness are
excluded from this minimum so that they do not drive $\Delta t_{\rm gas}\to0$.
These are flooring measures rather than a guarantee of strict monotonic
ordering: in a runaway cooling flow the affected shells would pile up at the floor
thickness and condense to the resolution limit, signaling the onset of a
cooling catastrophe, rather than physically crossing.

\section{An isolated cool-core halo}
\label{sec:application}

\subsection{Halo model and parameters}

To illustrate, we run a simulation where the dark matter follows a Hernquist profile~\cite{Hernquist:1990be},
\begin{equation}
     \rho_{\rm h}(r)=\frac{M_{\rm dm}}{2\pi}\frac{a_{\rm dm}}{r(r+a_{\rm dm})^3},
\end{equation}
and the gas follows a cored Dehnen~\cite{Dehnen:1993uh} profile with
$\gamma=0$,
\begin{equation}
     \rho_{\rm gas}(r)=\frac{3M_{\rm gas}}{4\pi}\frac{a_{\rm gas}}{(r+a_{\rm gas})^4},
\end{equation}
which is flat at small $r$ (a core) and falls as $r^{-4}$, qualitatively
matching the power-law-plus-core gas of relaxed cool-core
clusters~\cite{Vikhlinin:2006}.  The parameters correspond to a total mass
$M_{200}=10^{13}\,M_\odot$ within $r_{200}=454$~kpc at concentration
$c=6$~\cite{Diemer:2019}, a gas fraction equal to the cosmic value at
$r_{200}$ and rising to $0.75$ at $0.01\,r_{200}$, giving $M_{\rm dm}= 1.5\times10^{13}\,M_\odot$, $a_{\rm dm}=149$~kpc, $M_{\rm gas}=
1.6\times10^{12}\,M_\odot$, and $a_{\rm gas}=11$~kpc---a gas component that is centrally denser, and colder, than the dark matter. The Hernquist parameters follow from an NFW halo~\cite{Navarro:1996gj} with the same $r_{200}$ and $c$, rescaled to the dark-matter mass within $r_{200}$, using the central density-profile matching procedure of Ref.~\cite{Robertson:2016xjh}.
We take the dark-matter mass to be $m_\chi=m_H$ and a velocity-independent cross section per unit dark-matter mass of magnitude $\sigma_{\rm tot}/m_\chi=0.1\,{\rm cm}^2\,{\rm g}^{-1}$,  the CMB upper bound at this mass from Planck~\cite{Gluscevic:2017ywp}.

\begin{figure*}[!t]
\includegraphics[width=\textwidth]{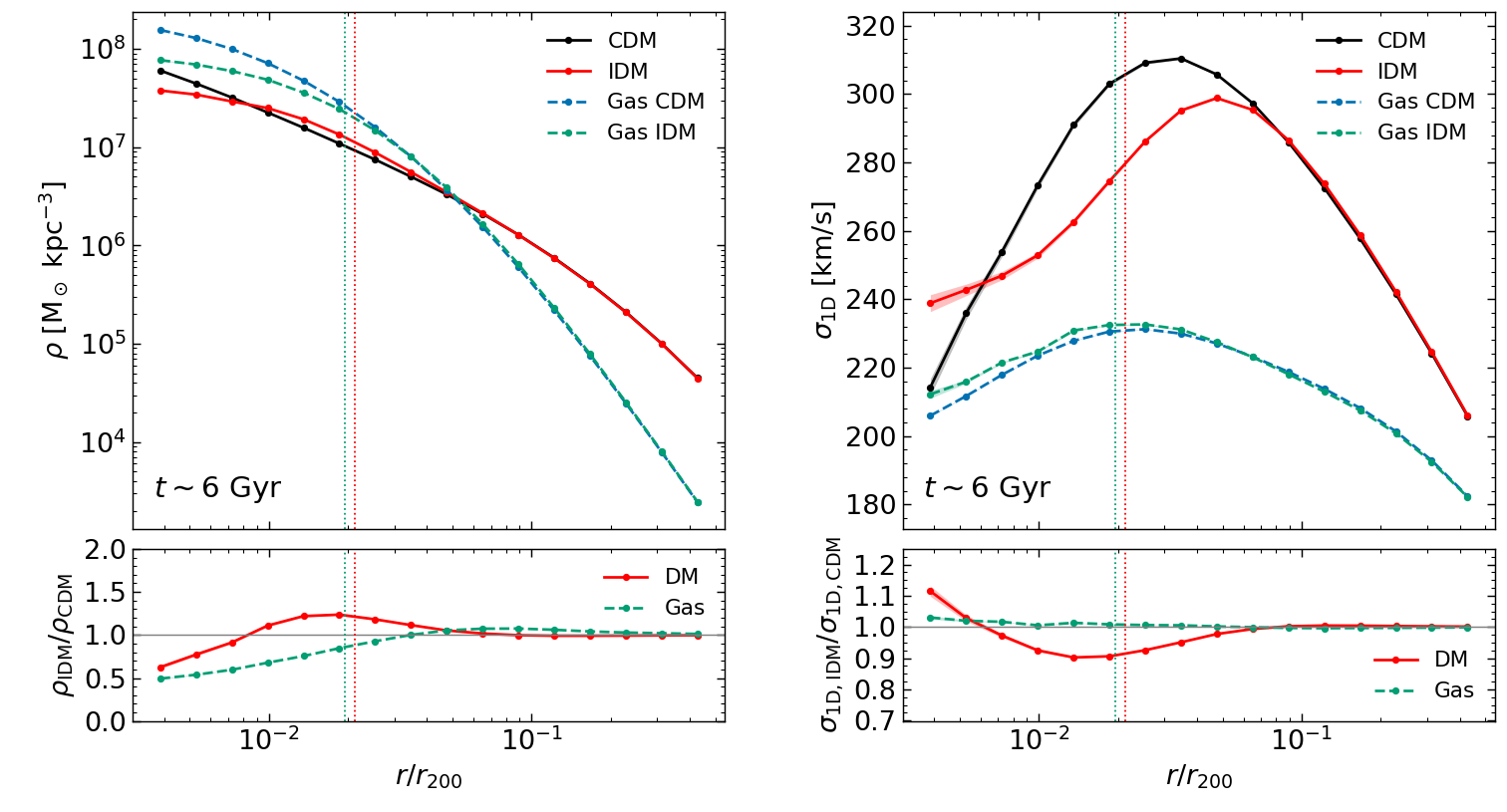}
\caption{Effect of dark-matter--baryon scattering, comparing the IDM simulation (red), with scattering on for the full $6$~Gyr, against the null run (CDM, black) at $t\sim6$~Gyr (three realizations, pooled over $5.8$--$6$~Gyr), as a function of $r/r_{200}$ (where here $r_{200}=454$ kpc).  {\it Left:} dark-matter (solid; CDM black, IDM red) and gas (dashed; CDM blue, IDM green) density, with the IDM/CDM ratios below (dark matter red, gas green). Scattering removes pressure-supported gas from the center, reducing the central gas density by  about a factor of two over the $6$~Gyr.  {\it Right:} one-dimensional velocity dispersion of the
dark matter (solid; CDM black, IDM red) and the gas thermal dispersion (dashed; CDM blue, IDM green), with the dark-matter and gas ratios below.  The dark matter is kinematically cooled, its dispersion reduced at radii of a few to a few tens of kiloparsecs by as much as $\sim10\%$. Vertical dotted lines mark the one-scattering radii (green for gas, red for dark matter), within which a typical gas atom ($r_{1,{\rm gas}}\simeq9$~kpc) or dark-matter particle ($r_{1,{\rm dm}}\simeq10$~kpc) scatters on average at least once over the 6~Gyr run. The shaded band on the IDM dark-matter dispersion at the smallest radii shows its $1\sigma$ sampling error, including the full covariance of the pooled particles. The bands on the other curves are mostly too narrow to be seen.}
\label{fig:cdmvsidm}
\end{figure*}

\subsection{Simulation}

We sample the dark matter with $N=2\times10^6$ numerical particles and the gas with $N_{\rm gas}=2000$ shells, out to $7.5\,r_{200}$.  We start our simulation with a dark-matter DF consistent with gas in hydrostatic equilibrium.  To isolate the effect
of scattering we run matched pairs of simulations---one with scattering, one without (the ``null'' run)---from identical initial conditions (the same random seed), and compare them after 6 Gyr.  We run three such pairs with different random seeds, and to reduce the sampling noise in the sparsely populated central bins we pool the particles of all three pairs over the last ten output snapshots ($5.8$--$6$~Gyr), over which the profiles appear stationary.  The run is split into 30,000 equally sized time steps.  The gas is evolved adiabatically (i.e., cooling is turned off).  A run with 1000 gas shells takes a few hours (wall time) for $10^6$ particles (and tens of minutes for $10^5$) on a 2022 MacBook Air with an M2 processor; over the $6$~Gyr, roughly $5.5\times10^4$ scattering events occur.

\subsection{Initial conditions}

Figure~\ref{fig:ics} shows the $t=0$ initial conditions.  The dark-matter density
tracks the analytic Hernquist profile---including the $r^{-1}$ cusp down to
$\sim2$~kpc---and the gas density tracks the analytic Dehnen $\gamma=0$
profile, across two decades in radius.  The dark-matter velocity dispersion matches the
isotropic-Jeans prediction, peaking near $310\,{\rm km\,s}^{-1}$ at
$\sim15$~kpc, and the gas thermal dispersion ($\sim200$--$230\,
{\rm km\,s}^{-1}$)  lies below it outside $\sim2$~kpc.  

\begin{figure*}[!t]
\includegraphics[width=\textwidth]{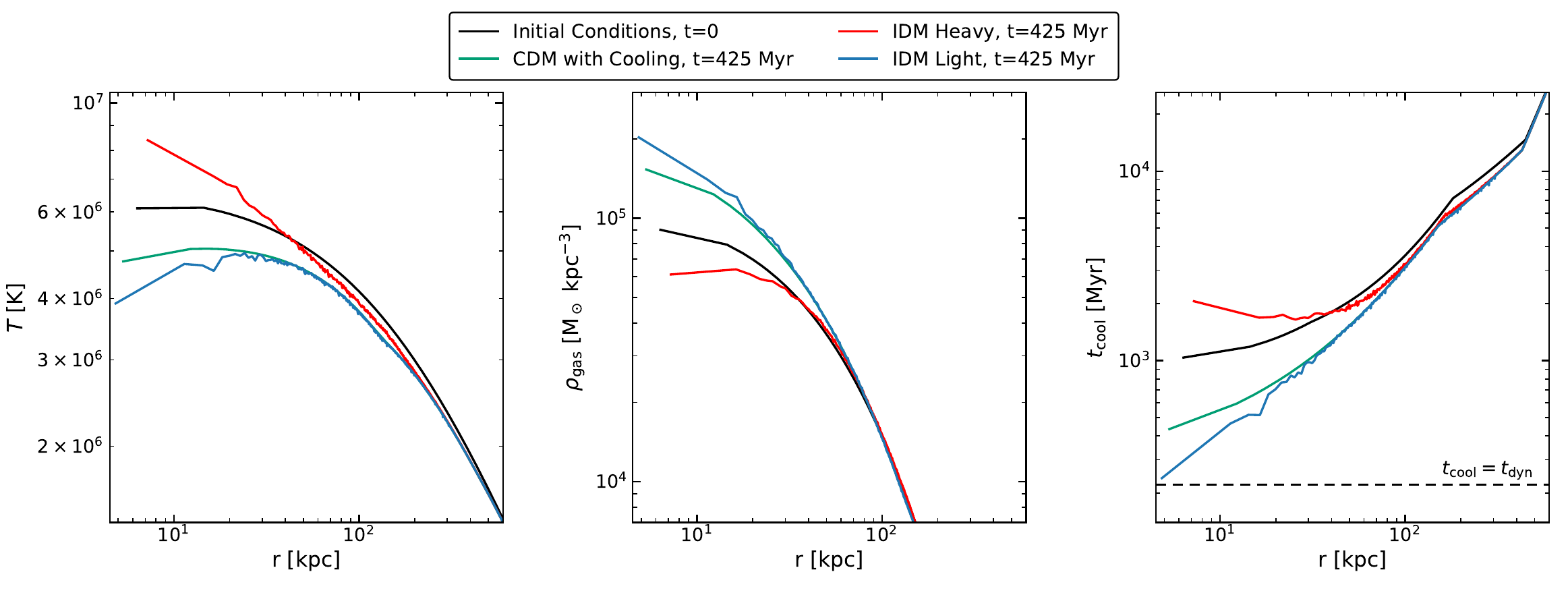}
\caption{Radiative cooling of gas  with dark-matter--baryon scattering. The dark-matter halo is initialized with a Hernquist profile
with mass $M_{\rm dm}$$=1.5\times10^{13}\,M_\odot$ and scale radius $a_{\rm dm}$$=149$~kpc, and the gas (at solar metallicity)
with a Dehnen profile of $M_{\rm gas}=1.6\times10^{12}\,M_\odot$ and scale radius $a_{\rm gas}=149$~kpc.  The dark matter is sampled with $10^{7}$ particles and the gas with $2000$ shells. The panels show, as a function of
radius, the gas temperature (left), density (center), and cooling time (right; the
dashed line marks the dynamical time $t_{\rm dyn}$), for the $t=0$ initial conditions  (black) and
for three evolved cases, including the cooling-only CDM simulation (green).  All three are shown at the same time, 425 Myr, just before the central cooling time of the light-dark-matter simulation (blue) falls to $t_{\rm dyn}$.  The red shows results with scattering from a dark-matter particle of mass $m_\chi=10\,m_H$ with cross section divided by particle mass of $\sigma_{\rm tot}/m_\chi=5\times10^{-24}\,{\rm cm}^{2}/m_H$ at the same time. 
The  blue curves show results with a dark-matter particle of mass $m_\chi=0.1\,m_H$ and the same $\sigma_{\rm tot}/m_\chi$, at the same time; the enhanced cooling due to the dark-matter--baryon interaction later leads to a cooling catastrophe (at $\approx550$~Myr, about 140 Myr before the cooling-only case). As expected, scattering from heavy (light) dark matter slows (accelerates) the cooling.}
\label{fig:cooling}
\end{figure*}

\subsection{Effects of scattering}

Figure~\ref{fig:cdmvsidm} compares the IDM run, with scattering acting over the full $6$~Gyr, to the null run at the same time.   Because the DF of Eq.~(\ref{eqn:DF}) with the density taken to be $\rho_{\rm dm}$ places the system in genuine equilibrium, the null run is stable---the gas density drifts by less than $2\%$ at all radii over $6$~Gyr, with no spurious feature near the gas scale radius, and the dark-matter cusp survives.

Comparing the null and IDM simulations, we see that a gas core
forms: the central gas density is reduced by  about a factor of two
(the IDM/CDM ratio reaches  $\simeq0.5$ at $r\approx2$~kpc), and the innermost
gas shell is pushed outward, as the energy and momentum deposited by the
dark matter remove pressure-supported gas from the center.  Second, the 
dark matter is cooled: its one-dimensional velocity dispersion is
reduced at radii of a few to a few tens of kiloparsecs by as much as $\sim10\%$ ($\sim30\,{\rm km\,s}^{-1}$), with an rms variation of $\sim8\%$ inside the peak of the CDM dispersion profile.  The dark-matter density is lowered at the center, by about 40\% at 1.75~kpc, and raised by up to 25\% at 6--10~kpc. We have checked that in both the null and IDM runs the dark-matter dispersion at $6$~Gyr satisfies the Jeans equation with the measured velocity anisotropy in the combined potential of the dark matter and gas, to within $\sim3\%$ (rms over 1--7~kpc).  The higher central dispersion of the IDM run, which stays flat while that of the CDM run falls toward the center, is consistent with the equilibrium response of a dark-matter core to the gas-dominated central potential.  Energy flows from the hotter, more diffuse dark matter into the colder, denser gas, heating and expanding the gas while cooling the dark matter.  The probability for a dark-matter particle to scatter over the duration of the simulation is estimated to be of order unity in the innermost 10 kpc, while the induced profile changes extend to radii of a few tens of kiloparsecs.

\section{Radiative cooling with dark-matter--baryon scattering}
\label{sec:coolingdemo}

Recent work \cite{Hu:2007ai,Shoji:2023pdg,Stuart:2024nkz} has studied the effects of dark-matter--baryon interactions on the cooling of gas in clusters, and there is analogous work on the gas evolution in dwarf galaxies \cite{Wadekar:2019mpc}.  That work used analytic models for the heating/cooling of gas, but it did not take into account the effects on the dark-matter distribution and thus is incomplete.  The numerical approach employed here should allow the complete physics to be modeled with modest computational cost.  We illustrate here with two simulations that include dark-matter--baryon interactions.

\subsection{Setup}

We keep the same Hernquist dark-matter halo as before, but choose the gas scale radius to coincide with that of the dark matter, so that the initial gas distribution resembles more closely that of the dark matter.  With this choice, the central cooling time is $\approx1040$~Myr, roughly 4.7 times the dynamical time (we define $t_{\rm dyn}=\sqrt{a_{\rm dm}^3/GM_{\rm dm}}\approx220$~Myr). Moreover, the central hydrostatic-equilibrium temperature 
is $\sim6\times10^{6}$~K, which sits past the metal-line peak of the cooling function, as expected for a halo of this mass at formation. Quoted temperatures in this section use the ionized mean molecular weight $\mu(x_e)\simeq0.6$ of collisional-ionization equilibrium; the pure-hydrogen value $\mu=1$ applies to the adiabatic runs. The thermal width entering the scattering rate, Eq.~(\ref{eqn:sigmav}), is evaluated at this same physical temperature.

We first run a cooling-only CDM simulation (cooling on, no scattering, solar metallicity) and two IDM simulations with the same initial conditions. The gas is sampled with 2000 shells.  The time step is taken to be $\Delta t\approx0.09$~Myr.  Our approach assumes that the gas remains close to hydrostatic equilibrium, which requires the cooling time to exceed the local free-fall time ($\approx40$~Myr in the core); we adopt the more conservative requirement that it exceed $t_{\rm dyn}$.  We therefore compare all three simulations at a common time,
$t_{\rm stop}\approx425$~Myr, just before the central cooling time of the light-dark-matter simulation, the first to cool, falls to $t_{\rm dyn}$ (at $\approx430$~Myr); at that time the cooling-only CDM case is still at about twice the dynamical time (it reaches $t_{\rm dyn}$ only at $\approx560$~Myr).  Up to that point the central gas cools gently
($T$ from $6.1$ to $4.8\times10^{6}$~K, $\rho_{\rm gas}$ rising by about 70\%) and only
afterward does the cooling time plunge.

We then run two IDM simulations, the first with $m_\chi=10\,m_H$ and another with $m_\chi=0.1\,m_H$.\footnote{ Given the scattering kinematics, the results should be fairly similar to the heavy-dark-matter simulation for any dark-matter particle mass $m_\chi \gg m_H$ with $\sigma_{\rm tot}/m_\chi$ fixed, and to the light-dark-matter simulation for any $m_\chi \ll m_H$ with $\sigma_{\rm tot}$ fixed.}    We choose the cross section (divided by dark-matter particle mass) for both to be $(\sigma_{\rm tot}/m_\chi)=5\times10^{-24}\,{\rm cm}^2/m_H$ so that the dark-matter--baryon heating/cooling rates are comparable in magnitude to the  radiative-cooling rate. This cross section is chosen for illustration; it exceeds current CMB bounds at these masses, by a factor of a few for the light-dark-matter run and by more than an order of magnitude for the heavy-dark-matter run.  All three simulations are compared at the same time, 425 Myr.  When the dark-matter mass is smaller than the hydrogen mass, dark-matter--baryon scattering transfers heat  from the gas to the dark matter.  With the given parameters, the central cooling time of the light-dark-matter simulation reaches $t_{\rm dyn}$ at $\approx430$~Myr, and its cooling catastrophe (the central gas reaching the $10^4$~K floor) occurs at $\approx550$~Myr, about 140 Myr earlier than in the cooling-only simulation.

\subsection{Results}

Figure~\ref{fig:cooling} compares the three simulations at 425 Myr.  In the cooling-only CDM simulation the gas behaves as expected for a cooling flow: the innermost shell cools
from $T=6.1$ to $4.8\times10^{6}$~K, its density rises by about 70\%, and its cooling time falls from $\approx1040$ to $\approx435$~Myr, on its way to catastrophic cooling.  In the heavy-dark-matter simulation heating from dark-matter--baryon interactions reverses this: the central gas is heated to $T\approx8.4\times10^{6}$~K, expands so that its central density drops by about a third, and its cooling time rises to $\approx2.1\times10^{3}$~Myr, far greater than the dynamical time.  Thus heating by dark matter more
than offsets radiative losses and prevents runaway cooling.  When the dark matter is lighter than hydrogen, dark-matter--baryon scattering cools the gas, thus accelerating the cooling catastrophe.  The noise in the  blue and red curves  arises from Poisson fluctuations in the scattering Monte Carlo.  These fluctuations are larger in the light-dark-matter simulation than in the heavy-dark-matter simulation because the cross section in the light-dark-matter simulation is smaller (although the dark matter-particle mass is different in the two simulations, the numerical particle masses are the same).  At the same time the light-dark-matter gas (blue) is colder and denser than the cooling-only CDM gas (green), with a central cooling time of $\approx240$~Myr, already close to $t_{\rm dyn}$, against $\approx435$~Myr.

\section{Conclusions}
\label{sec:conclusions}

We have presented a fast method for evolving a spherically symmetric halo
of collisionless dark matter and gas, built on the
spherical $N$-body code {\tt NSphere}.  We have also implemented dark-matter--baryon scattering and included radiative cooling.  We showed how to set initial conditions for a static configuration of dark matter and gas in hydrostatic equilibrium.  The gas is evolved as Lagrangian shells, and scattering is implemented as a Monte-Carlo process that  conserves energy and radial momentum exactly in each event, depositing both heat and radial momentum into the gas.  We demonstrated the method by modeling the evolution of a cluster-mass dark-matter halo with a cool gas core first in a simulation with adiabatic gas evolution (no cooling).  We then illustrated this with a simulation including radiative cooling.  In a diffuse gas that would otherwise cool catastrophically, scattering by heavy dark matter ($m_\chi=10\,m_H$) heats the gas and prevents the cooling catastrophe.  A simulation with light dark matter ($m_\chi=m_H/10$) accelerates the cooling.

The limitations of our approach are obvious: real galaxies are not perfectly round, nor are they formed from spherically-symmetric initial conditions.  Thus, our code can never replace full  3D simulations.  We have also not yet included  shocks and so are restricted to situations where cooling times are long compared with dynamical times and also to gas flows that remain subsonic.

Still, dark-matter models span many values of dark-matter mass, dark-matter--baryon cross section, and self-interaction cross sections.  There  is a plethora of different types of velocity and angular dependences of those dark-matter--baryon cross sections.  And there are uncertainties in the velocity distributions of dark matter.  A fast robust numerical calculation is thus valuable for getting intuition and  estimates for broad swaths of dark-matter parameter space.  The approach will also be useful to cross-check and calibrate full  3D simulations, which often need to make simplifying assumptions or approximations about gravitational dynamics (e.g., gravitational softening) or the dark-matter--baryon interaction.

Here we considered, by way of illustration, only velocity-independent and isotropic scattering cross sections, but generalization to other types of cross sections, and to dark-matter particles that scatter from electrons, rather than protons, is straightforward.  The combined effects of dark-matter self-interactions and dark-matter--baryon scattering can be investigated.   We initialized the dark matter with an isotropic velocity distribution, but the effects of anisotropy can be explored.

There are any number of elaborations and expansions that should be simple to implement with little extra computational cost.  For the purposes of illustration, we made some simplifying assumptions---e.g., that dark-matter particles scatter as if all the baryons were free protons and the scattering rate uses the total baryon number density, $n=\rho_{\rm gas}/m_H$ (Sec.~\ref{sec:scattering}). Short of implementing the exact per-species kinematics, the effective cross section of Appendix~\ref{app:sigmaeff} captures the composition dependence and provides a quantitative mapping of our results (and {\tt NSphere-IDM} more generally) onto alternative microphysical models. We did not check whether gas in near hydrostatic equilibrium satisfies the Schwarzschild criterion, and this, as well as a prescription for convection, can be added.
Given our focus on galaxy-size halos, which tend to be optically  thin, we did not include radiative transfer, but it should be straightforward to do so.  It should be possible to develop a prescription for the evolution into the catastrophic cooling regime and also one for star formation.  Stellar orbits are also easily added.  By considering gas at much higher densities and adding nuclear energy generation (and with radiative transport), the approach can be used to model dark stars \cite{Freese:2010re}.

\begin{acknowledgments}
We thank James Wen, Ethan Nadler, and Moritz Fischer for making available the details of the simulations of Ref.~\cite{Wen:2026}.  MK was supported by NSF Grant No.\ 2412361, NASA ATP Grant No.\ 80NSSC24K1226, and the John Templeton Foundation Grant No.\ 62840.  VG was supported by the John Templeton Foundation Grant No.\ 63667, the Research Corporation for Science Advancement Cottrell Scholar Program, the IBM Einstein Fellowship at the Institute for Advanced Study, and the NSF CAREER Grant No. PHY-2239205. KS was supported by a Discovery Grant from the Natural Sciences and Engineering Research Council of Canada.
\end{acknowledgments}

\appendix

\section{Numerical implementation of the gas evolution}
\label{app:gas}

Here we describe how we discretize and advance the gas equations of Sec.~\ref{sec:gas}. The gas is represented by $N_{\rm gas}$ concentric spherical shells bounded by $N_{\rm gas}+1$ edges at radii $r_i$ ($i=0,1,\dots,N_{\rm gas}$), with $r_0=0$. The evolved variables are the edge radii $r_i$, the edge (radial) velocities $v_i$, and the shell-centered specific internal energies $u_{i+\frac12}$.  The shell density $\rho_{i+\frac12}$ and pressure $P_{i+\frac12}$ are derived from these.

\subsection{Shell edges and initial conditions}

The shells carry equal gas mass, $\Delta m_{\rm gas}\equiv M_{\rm gas}/N_{\rm gas}$, so the edge
$r_i$ encloses gas mass $m_i=i\,\Delta m_{\rm gas}$. The edge radii are set by analytically
inverting the cumulative-mass profile $M_{\rm gas}(<r)=M_{\rm gas}\,f$ at
$f=m_i/M_{\rm gas}$. For the two gas profiles  implemented in {\tt NSphere-IDM} this gives, respectively,
\begin{eqnarray}
r_i &=& a_{\rm gas}\,\frac{f^{1/3}}{\sqrt{1-f^{2/3}}}\quad(\text{Plummer}), \\
r_i &=& a_{\rm gas}\,\frac{f^{1/3}}{1-f^{1/3}}\quad(\text{Dehnen }\gamma{=}0).
\label{eqn:edgeinvert}
\end{eqnarray}
For the $r^{-4}$ (Dehnen) tail the mass converges only slowly, so the edges are
distributed over $f\in[0,f_{\max}]$ with $f_{\max}=\min[M_{\rm gas}(<r_{\rm cut})/M_{\rm gas},\,0.99]$
and $r_{\rm cut}$ the sampling radius;  each shell then carries mass $f_{\max}\Delta m_{\rm gas}$ (and $\Delta m_{\rm gas}$ is replaced by $f_{\max}\Delta m_{\rm gas}$ in what follows). The shell density follows from the edge radii,
\begin{equation}
\rho_{i+\frac12} = \frac{\Delta m_{\rm gas}}{\tfrac{4}{3}\pi\,(r_{i+1}^3-r_i^3)},
\label{eqn:shellrho}
\end{equation}
and the equation of state $P_{i+\frac12}=\tfrac23\rho_{i+\frac12}u_{i+\frac12}$ closes the system
with $u=(3/2) k_BT(\mu m_H)^{-1}$. The initial $u_{i+\frac12}$ is obtained by imposing
hydrostatic equilibrium: the pressure $P(r)=\int_r^\infty G\,M_{\rm tot}(<r')\,\rho_{\rm gas}(r')/r'^2\,dr'$
is computed once on a fine logarithmic grid (right-to-left cumulative trapezoidal sum,
with $P=0$ well outside the gas surface) and interpolated to the shell centers, whence
 $u_{i+\frac12}=(3/2)\,P_{i+\frac12}/\rho_{\rm gas}(\bar r_{i+\frac12})$, with the profile density evaluated at the shell midpoint $\bar r_{i+\frac12}=(r_i+r_{i+1})/2$. The edge velocities are initialized to zero.

\subsection{Discrete equations of motion}

Written as first-order equations for the edges, $\dot r_i=v_i$ and $\dot v_i=a_i$, where the dot denotes derivative with respect to time, the
acceleration is the discretization of Eq.~(\ref{eqn:fluidacceleration}),
\begin{equation}
a_i = -\,\frac{G\,M_{\rm tot}(<r_i)}{r_i^2}
      \;-\;4\pi r_i^2\,\frac{P_{i+\frac12}-P_{i-\frac12}}{\Delta m_{\rm gas}}
      \;+\;\bar F_{\chi b,\,i}.
\label{eqn:accel}
\end{equation}
The first term is gravity: $M_{\rm tot}(<r_i)=M_{\rm dm}(<r_i)+i\,\Delta m_{\rm gas}$, where the
dark-matter contribution is read off the radius-sorted collisionless particles (the same
enclosed mass used for their own forces), and $i\,\Delta m_{\rm gas}$ is the gas interior to edge
$i$. The second term is the pressure gradient $-4\pi r^2\,\partial P/\partial m$,
evaluated as a centered difference of the two shells adjoining edge $i$.  We impose as a boundary condition that the pressure at the surface is
$P_{N_{\rm gas}+1/2}\equiv0$ (free boundary). The last term in Eq.~(\ref{eqn:accel}) is the
dark-matter--baryon scattering force: the shell-centered accelerations $F_{\chi b,i+\frac12}$
are averaged to the edge, $\bar F_{\chi b,i}=(1/2)(F_{\chi b,i-1/2}+F_{\chi b,i+1/2})$. The center is held
fixed and regular, with $r_0=0$, $v_0=0$, and $a_0=0$.

\subsection{Kick--drift--kick update}

Over a step $\Delta t$ the mechanical variables are advanced with a velocity leapfrog,
while the thermal state is updated during the drift:
\begin{align}
&\text{(K)}\quad v_i^{\,n+\frac12}=v_i^{\,n}+\tfrac12 a_i^{\,n}\,\Delta t,\nonumber\\
&\text{(D)}\quad r_i^{\,n+1}=r_i^{\,n}+v_i^{\,n+\frac12}\,\Delta t,\nonumber\\
&\phantom{\text{(D)}}\quad \rho_{i+\frac12}^{\,n+1}\ \text{from Eq.~(\ref{eqn:shellrho})};\nonumber \\
    & \qquad
   u_{i+\frac12}\leftarrow u_{i+\frac12}\!\left(\frac{\rho_{i+\frac12}^{\,n+1}}{\rho_{i+\frac12}^{\,n}}\right)^{2/3}\!\!+\,\epsilon_{\chi b}\,\Delta t,\nonumber\\
&\text{(K)}\quad v_i^{\,n+1}=v_i^{\,n+\frac12}+\tfrac12 a_i^{\,n+1}\,\Delta t.
\label{eqn:kdk}
\end{align}
The accelerations $a_i^{\,n}$ and $a_i^{\,n+1}$ are evaluated from
Eq.~(\ref{eqn:accel}) before and after the drift. In the drift, the density is recomputed
from the new edges; the specific internal energy is advected adiabatically
($u\,\rho_{\rm gas}^{-2/3}={\rm const}$, i.e.\ $P\propto\rho_{\rm gas}^{5/3}$); the scattering heat
$\epsilon_{\chi b}$ of Eq.~(\ref{eqn:deposit}) is added; and $P_{i+\frac12}$ is refreshed
from the equation of state. (In the code, velocities are stored in ${\rm kpc/Myr}$ while
$u$ is in $({\rm km/s})^2$; the pressure term of Eq.~(\ref{eqn:accel}) carries the corresponding unit conversion.)

\subsection{Source terms, subcycling, and shell ordering}

A fully complete treatment that includes radiative cooling would require substantial augmentations to the code.  Here we incorporate radiative cooling using a scheme that should approximate the evolution in situations where the cooling time is not shorter than the dynamical time---we leave a fuller development to future work.

In our calculation, radiative cooling (Sec.~\ref{sec:cooling}) is evolved separately from the dynamics: after the adiabatic-plus-scattering update of Eq.~(\ref{eqn:kdk}), the internal energy of each shell is evolved under cooling alone at fixed density, using Townsend integration \cite{Townsend:2009mf}. Because rapid cooling steepens the pressure gradients, the global step is subcycled with a Courant-limited substep, $\Delta t_{\rm gas}=C_{\rm cfl}\,\min_i[\Delta r_i/(c_{s,i}+|v_i|)]$ with $C_{\rm cfl}=0.3$ and $c_s=\sqrt{(5/3)P/\rho_{\rm gas}}$, so an edge advances by only a fraction of its shell width per substep. Finally, to keep the Lagrangian shells well ordered, after each drift the edges are clamped onto a strictly increasing grid: $r_0\ge0$ and $r_i\ge r_{i-1}+\delta r$ with a floor $\delta r=10^{-4}a_{\rm gas}$; an edge violating this
(or returning a non-finite value) is reset to $r_{i-1}+\delta r$ and its inward velocity is zeroed. Shells thinner than twice the floor, $\Delta r_i\le2\,\delta r$, are excluded from the Courant minimum so that a single collapsed shell cannot drive $\Delta t_{\rm gas}\to0$. These flooring measures keep the grid ordered but do not guarantee a physical evolution: in a runaway cooling flow the affected shells pile up at the floor thickness rather than physically cross, signaling the onset of a cooling catastrophe.

\section{\texorpdfstring{Effective cross section for a hydrogen--helium or metal-rich gas}{Effective cross section for a hydrogen-helium or metal-rich gas}}
\label{app:sigmaeff}

The scattering model of Sec.~\ref{sec:scattering} treats the gas as free nucleons: the targets carry the total nucleon number density $n=\rho_{\rm gas}/m_H$, a single cross section $\sigma_{\rm tot}$ per nucleon, and recoil kinematics with target mass $m_H$. Here we relate this model to coherent spin-independent couplings $f_p$ and $f_n$ to protons and neutrons, for a hydrogen--helium gas with mass fractions $X=0.71$ and $Y=0.27$.

At the velocities and dark-matter masses of interest the momentum transfer is $q\sim0.1$--$10$~MeV, far below the inverse nuclear size of helium, $\hbar c/R_{\rm He}\simeq120$~MeV, so scattering off helium is always coherent. At these same momentum transfers scattering is fully incoherent between atoms, which therefore scatter independently even in molecular gas. The helium cross section is $\sigma_{\rm He}=[Z+(f_n/f_p)(A-Z)]^2(\mu_{\chi{\rm He}}/\mu_{\chi p})^2\,\sigma_p$, where $\sigma_p$ is the dark-matter--proton cross section and $\mu_{\chi{\rm He}}$ and $\mu_{\chi p}$ are the dark-matter--helium and dark-matter--proton reduced masses~\cite{Kurylov:2003ra}. The coherence factor is $A^2=16$ for equal couplings, $Z^2=4$ for proton-only couplings, and zero for $f_n=-f_p$, for which the helium ($A=2Z$) amplitude cancels~\cite{Feng:2011vu}. Helium also decouples for spin-dependent couplings, since $^4$He has spin zero.

The gas evolution in this work is driven by the volumetric dark-matter--gas energy-exchange rate. The contribution of nuclear species $i$ is proportional to $n_i\sigma_i\kappa_i\langle v_{\rm rel}\rangle_i$, where $n_i=X_i\rho_{\rm gas}/(A_i m_H)$ is its number density and $\kappa_i=m_\chi m_i/(m_\chi+m_i)^2$ sets the energy exchanged per collision~\cite{Dvorkin:2013cea,Munoz:2015bca}. Matching the total exchange rate of the hydrogen--helium mixture to that of the free-nucleon model gives the effective cross section
\begin{equation}
\sigma_{\rm tot}^{\rm eff}=\sigma_p\left\{X+\frac{Y}{4}\,\eta_{\rm He}\left(1+\frac{f_n}{f_p}\right)^{\!2}\left[\frac{4(m_\chi+m_H)}{m_\chi+4m_H}\right]^{4}\right\},
\label{eqn:sigmaeff}
\end{equation}
in which the fourth power of the reduced-mass ratio combines the coherent enhancement of the cross section ($\mu^2$) with the per-collision energy transfer ($\mu^2/m_i$), and $\eta_{\rm He}\le1$ is a thermal-width factor defined below. For a general composition,
\begin{equation}
\sigma_{\rm tot}^{\rm eff}=\sigma_p\sum_i\frac{X_i}{A_i^2}\,\eta_i\,\mathcal{F}_i\left[Z_i+\frac{f_n}{f_p}\,(A_i-Z_i)\right]^2\left(\frac{\mu_{\chi i}}{\mu_{\chi p}}\right)^{\!4},
\label{eqn:sigmaeffgen}
\end{equation}
where the sum runs over nuclear species $i$ with mass number $A_i$, atomic number $Z_i$, and mass fraction $X_i$. The factor $\mathcal{F}_i\le1$ is the energy-transfer-weighted nuclear form-factor suppression~\cite{Helm:1956zz,Lewin:1995rx}; it is unity to excellent accuracy for hydrogen and helium, and the hydrogen term reduces to $X\sigma_p$.\footnote{These expressions describe the energy-exchange channel that drives the effects studied here; the momentum-transfer (drag) channel~\cite{Dvorkin:2013cea} weights species as $(X_i/A_i)[Z_i+(f_n/f_p)(A_i-Z_i)]^2(\mu_{\chi i}/\mu_{\chi p})^3$ and is negligible for the subsonic flows we consider.}

The factor $\eta_{\rm He}$ accounts for the difference between the hydrogen and helium thermal widths: for Maxwellian velocities the mean relative speed for species $i$ is $\propto(\sigma_v^2+k_BT/m_i)^{1/2}$, where $\sigma_v$ is the one-dimensional dark-matter velocity dispersion, so
\begin{equation}
\eta_{\rm He}=\left[\frac{\sigma_v^2+k_BT/m_{\rm He}}{\sigma_v^2+k_BT/m_H}\right]^{1/2}.
\label{eqn:etaHe}
\end{equation}
It lies between $1/2$ (gas-thermal-dominated) and unity (dispersion-dominated) and is $\simeq0.85$ for the conditions simulated here ($\sigma_v\simeq280\,{\rm km\,s^{-1}}$, $k_BT/m_H\simeq(220\,{\rm km\,s^{-1}})^2$), falling to $\simeq0.8$ at the center, where $\sigma_v$ approaches the gas thermal speed. In Eq.~(\ref{eqn:sigmaeffgen}), $\eta_i$ is the same expression with $m_{\rm He}\rightarrow m_i$.

Three limits of Eq.~(\ref{eqn:sigmaeff}) are useful. For $m_\chi\ll m_H$ the reduced-mass ratio tends to unity, so for equal couplings $\sigma_{\rm tot}^{\rm eff}\to(X+\eta_{\rm He}Y)\,\sigma_p\simeq\sigma_p$ and the free-nucleon model reproduces the energy-exchange rate to within $\sim5$--$10\%$. The $A^2$ enhancement on helium is offset by the smaller number of nuclei per unit mass and the smaller energy transfer per collision on the heavier nucleus. For $m_\chi\gg m_H$ the ratio tends to 4: the equal-coupling effective cross section reaches $\simeq27\,\sigma_p$ at $m_\chi=10\,m_H$ and saturates at $\simeq70\,\sigma_p$ for $\eta_{\rm He}\rightarrow1$ ($\simeq23$ and $\simeq60\,\sigma_p$ at $\eta_{\rm He}=0.85$). At fixed $\sigma_p$, coherent scattering on helium thus enhances the energy-exchange rate by more than an order of magnitude. For $f_n=-f_p$ helium drops out, and the free-nucleon model is exactly equivalent, for any $m_\chi$, to scattering off hydrogen alone with $\sigma_{\rm tot}=X\sigma_p\simeq0.71\,\sigma_p$.

Equation~(\ref{eqn:sigmaeff}) applies to gas without metals. At the solar metallicity adopted for the cooling calculations of Sec.~\ref{sec:coolingdemo}, the metals enter through the terms of Eq.~(\ref{eqn:sigmaeffgen}) with $Z_i>2$, whose sum we denote $\sigma_{\rm met}^{\rm eff}$. For the heaviest nuclei the momentum transfer approaches the inverse nuclear size, and $\mathcal{F}_i$ falls below unity. For $m_\chi\lesssim m_H$ the metals are a small correction: $\simeq10\%$ of $\sigma_{\rm tot}^{\rm eff}$ at $m_\chi=m_H$, approaching the $\simeq2\%$ metal mass fraction as $m_\chi\rightarrow0$, where, for equal couplings, Eq.~(\ref{eqn:sigmaeffgen}) becomes nearly composition-independent. For equal couplings the metals, though only $\sim2\%$ by mass, contribute about twice as much as helium at $m_\chi=10\,m_H$ and increasingly dominate at larger $m_\chi$. For $f_n=-f_p$ the neutron-excess isotopes that survive the cancellation (chiefly $^{56}$Fe) contribute increasingly with $m_\chi$: the hydrogen-only equivalence holds to better than $1\%$ for $m_\chi\lesssim5\,m_H$ and to $10\%$ for $m_\chi\lesssim10\,m_H$. Near $m_\chi\simeq18\,m_H$, $\sigma_{\rm met}^{\rm eff}=(1-X)\,\sigma_p\simeq0.29\,\sigma_p$, so that $\sigma_{\rm tot}^{\rm eff}=\sigma_p$, and beyond this it continues to grow.

Since $\sigma_{\rm tot}$ is illustrative in this work and the composition is taken to be uniform, these composition effects amount to a reinterpretation of the cross section rather than a qualitative change in the dynamics. The mapping is quantitative throughout the regime simulated here. It degrades for cross sections large enough that each dark-matter particle scatters several times before the dark matter reaches the gas temperature. The dark-matter velocity dispersion then changes appreciably, so $\eta_{\rm He}$ and $\sigma_{\rm tot}^{\rm eff}$ change too. This degradation is transient, since once the dark matter has thermalized with the gas its velocity distribution is the same for any gas composition. The mapping also degrades (i) for contact interactions with $\sigma_p$ approaching the geometric nuclear cross section, where the coherent Born scaling breaks down~\cite{Digman:2019wdm}; (ii) for relative velocities $\gtrsim10^{-2}c$, where the helium form factor enters and, for $m_\chi\gtrsim10\,m_H$, inelastic scattering on iron opens; and (iii) for velocity-dependent or anisotropic cross sections, for which the coherence and kinematic factors no longer combine into a single universal form. Abundance gradients, such as helium sedimentation or centrally concentrated metals from enrichment and mass segregation, add a radial dependence to $\sigma_{\rm tot}^{\rm eff}$ beyond that from the gas temperature and dark-matter dispersion profiles. The effect enters only through the species dependence of the kinematic factors. It is largest for $m_\chi\gtrsim10\,m_H$, where helium and metals dominate, and, for equal couplings, becomes very small for $m_\chi\ll m_H$.

\bibliography{refs}

\end{document}